\documentclass[english]{article}
\usepackage[T1]{fontenc}
\usepackage[latin1]{inputenc}
\usepackage{geometry}
\usepackage{graphicx}
\usepackage{amssymb}

\makeatletter

\providecommand{\tabularnewline}{\\}

\usepackage{babel}
\makeatother
\begin{document}

\title{\textbf{Branching ratios of radiative transitions in O VI}}

\author{Chiranjib Sur \\
 \emph{Department of Astronomy, Ohio State University, Columbus, Ohio,
43210, USA}\\
\\
Rajat K Chaudhuri \emph{}\\
\emph{Indian Institute of Astrophysics, Koramangala, Bangalore, 560
034, India }}

\date{(J. Phys. B : Communicated March 30, 2007)}

\maketitle
\begin{abstract}
We study the branching ratios of the allowed and forbidden radiative
transitions among the first few (9) fine structure levels of O VI
using relativistic coupled cluster theory. We find irregular patterns
for a number of transitions with in $n$-complexes with $n\le4$.
We have used the exisiting values of the allowed electric dipole ($E1$)
transition as a benchmark of our theory. Good agreement with the existing
values establish accuracies of not only the theoretical method but
the basis function as well. In general the electric quadrupole ($E2$)
transition probabilities are greater in magnitude than magnetic dipole
($M1$) transition probabilities, whereas for medium atomic transition
frequencies they are of the same order of magnitude. On the other
hand if the  transitions involved are between two fine structure components
of the same term, then the $M1$ transition probability is more probable
than that of $E2$. We have analyzed these trends with physical arguments
and order of magnitude estimations. The results presented here in
tabular and graphical forms are compared with the available theoretical
and observed data. Graphical analysis helps to understand the trends
of electric and magnetic transitions for the decay channels presented
here. Our calculated values of the lifetimes of the excited states
are in very good agreement with the available results. 

~

\end{abstract}

\section{Introduction}

Comprehensive and accurate transition probability data sets are needed
to determine the abundances of neutral oxygen and all of its ions
in different astrophysical spectra. Determination of radiative lifetimes
can provide the absolute scale for converting the branching fractions
into atomic transition probabilities and vice-versa. The emission
line diagnostics of allowed and forbidden transitions for O VI have
been used for a wide range of spectral modeling of planetary nebulae
and H II regions \cite{feibelman-1,feibelman-2} and in NGC 2867 \cite{feibelman-3}.
The range of these spectral studies lies between ultraviolet (UV)
and optical region. The most prominent emission line doublets of O
VI are ($\lambda\lambda$ 1032, 1038) and ($\lambda\lambda$ 3811,
3834). They lie in the far-UV and optical regions of the spectra respectively
in many central stars. Moreover O VI is one of the highest observed
states of ionization in stellar spectra and is thus considered a rich
source of data for high temperature plasma diagnostics. Tayal \cite{tayal}
and Aggarwal \emph{et al} \cite{aggarwal} provide more references.
Tayal \cite{tayal} has studied the allowed transitions and electron
impact collision strengths in the Breit-Pauli R-matrix framework whereas
Aggarwal \emph{et al.} \cite{aggarwal} have employed the fully relativistic
GRASP \cite{grasp} and DARC \cite{darc} codes for the same study.
Froese Fischer \emph{et al.} have employed multi-configuration Hartree-Fock
(MCHF) method to calculate the allowed ($E1$) transition probabilities
and the lifetime of the excited states \cite{mchf}.

The calculation of radiative transitions for many-electron atoms requires
a good description of the electron correlation which determines the
movement of the radiating electrons. In this article we have used
relativistic open-shell coupled cluster method (RCCM), one of the
most accurate theories to describe the electron correlation effects
in many-electron atoms, to calculate allowed and forbidden radiative
transition probabilities. 

This article deals with the study of allowed and forbidden transitions
in O VI ion using relativistic coupled cluster (RCC) theory. The present
study of the allowed and forbidden transitions is used to identify
the dependencies of the line strengths, transition probabilities and
oscillator strengths on the transition energies. From the definition,
the dependencies of transition probabilities and oscillator strengths
on the transition energies is quite clear but we have observed that
the line strengths also follow a trend which can be understood by
studying the physics involved in the problem. Graphical analysis of
the results presented in this article can be used to understand the
strengths of different electromagnetic transitions for a particular
emission.

\begin{figure}
\begin{centering}
\includegraphics[scale=0.75]{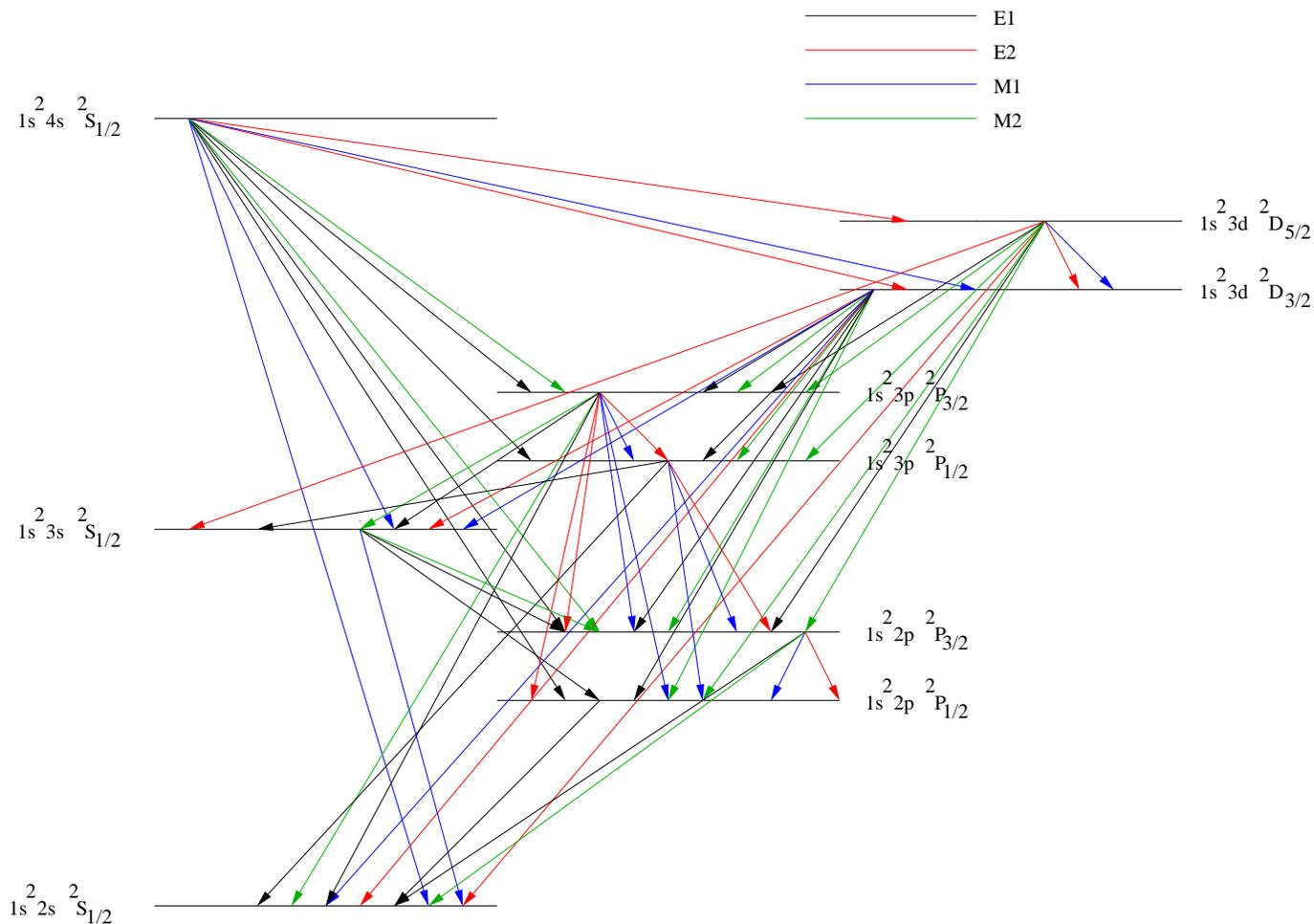}
\par\end{centering}

\caption{\label{decay-channels}Decay channels for the first few excited states
of O VI. The lines with different colours correspond to different
electromagnetic (multipole) transitions.}
\end{figure}

\section{Theoretical methods}

In this work we have employed RCCM to study the branching ratios of
O VI. The virtue of RCCM is that it is an all-order non-perturbative
theory and size extensive in nature \cite{size-extensive}. Higher
order electron correlation effects can be incorporated more efficiently
than by using order-by-order diagrammatic many-body perturbation theory
(MBPT). Extensive literature is available on RCC theory and application
to calculate atomic properties \cite{kaldor-book,lindgren-book},
so it will not be provided here. Our method of calculation is described
in details in a recent review article \cite{napp-review}.

\subsection{Radiative transitions}

The interaction of electromagnetic radiation with atoms in most plasma
applications is generally treated in what is known as \emph{first
quantization}, \emph{i.e.} the quantized energy levels of the atom
interact with a continuous radiation field. For spontaneous decay,
transition probabilities ($\mathtt{A}$) of different electromagnetic
multipoles are well known. Oscillator strengths ($\mathtt{f}$) and
hence the weighted oscillator strengths ($g\mathtt{f}$), which are
the products of the degeneracies of the initial states of the transition
and $\mathtt{f}$ are very important in astrophysics and can be used
to obtain information about the spectral formation and other quantities
of astrophysical interests. The knowledge of transition probabilities
is often used to determine the line ratios of the density diagnostics
of the plasmas. 

In this article $Ek$($Mk$) stands for electric (magnetic) type transitions
and $k=1,2..$ correspond to dipole, quadrupole transition respectively.
Here $\mathtt{S}^{Ek/Mk}$ are the line strengths; $\lambda$ is the
wavelength in $\textrm{\AA}$ and $\omega_{f}$ is the statistical
weight for the upper level for a transition $f\longrightarrow i$.
$\mathtt{A}_{fi}$ and $\mathtt{f}_{fi}$ are the transition probabilities
and oscillator strengths respectively. The $\mathtt{A}$ and $\mathtt{f}$
values for different types of transitions with the line strength ($\mathtt{S}$)
are related by the following standard equations :\\
 for the electric dipole ($E1$) transitions : 

\begin{equation}
\mathtt{A}_{fi}=\frac{2.0261\times10^{18}}{\omega_{f}\lambda_{fi}^{3}}\mathtt{S}^{E1}\,\,\,\mathrm{and}\,\,\,\mathtt{f}_{fi}=\frac{303.75}{\lambda_{fi}\omega_{i}}\mathtt{S}^{E1},\label{E1}\end{equation}
 for the electric quadrupole ($E2$) transitions : 

\begin{equation}
\mathtt{A}_{fi}=\frac{1.1199\times10^{18}}{\omega_{f}\lambda_{fi}^{5}}\mathtt{S}^{E2}\mathrm{\,\, and\,\,}\,\mathtt{f}_{fi}=\frac{167.89}{\lambda_{fi}^{3}\omega_{i}}\mathtt{S}^{E2},\label{E2}\end{equation}
for the magnetic dipole ($M1$) transitions : 

\begin{equation}
\mathtt{A}_{fi}=\frac{2.6974\times10^{13}}{\omega_{f}\lambda_{fi}^{3}}\mathtt{S}^{M1}\,\,\mathrm{and}\,\,\mathtt{f}_{fi}=\frac{4.044\times10^{-3}}{\lambda_{fi}\omega_{i}}\mathtt{S}^{M1}\label{M1}\end{equation}
and for the magnetic quadrupole ($M2$) transitions : 

\begin{equation}
\mathtt{A}_{fi}=\frac{1.4910\times10^{13}}{\omega_{f}\lambda_{fi}^{5}}\mathtt{S}^{M2}\,\,\mathrm{and}\,\,\mathtt{f}_{fi}=\frac{2.236\times10^{-3}}{\lambda_{fi}^{3}\omega_{i}}\mathtt{S}^{M2}\,.\label{M2}\end{equation}
The single particle reduced matrix elements for the $E1$, $E2$,
$M1$ and $M2$ transitions are given by,

\begin{equation}
\left\langle \kappa_{f}\right\Vert e1\left\Vert \kappa_{i}\right\rangle =\left\langle j_{f}\right\Vert C_{q}^{(1)}\left\Vert j_{i}\right\rangle \times\int r\left(P_{f}P_{i}+Q_{f}Q_{i}\right)dr\label{e1-mat}\end{equation}
\begin{equation}
\left\langle \kappa_{f}\right\Vert e2\left\Vert \kappa_{i}\right\rangle =\frac{15}{k^{2}}\left\langle j_{f}\right\Vert C_{q}^{(2)}\left\Vert j_{i}\right\rangle \times\int\mathtt{j}_{2}(kr)\left(P_{f}P_{i}+Q_{f}Q_{i}\right)dr,\label{e2-mat}\end{equation}
\begin{equation}
\left\langle \kappa_{f}\right\Vert m1\left\Vert \kappa_{i}\right\rangle =\frac{6}{\alpha k}\left\langle j_{f}\right\Vert C_{q}^{(1)}\left\Vert j_{i}\right\rangle \times\int\frac{\kappa_{f}+\kappa_{i}}{2}\mathtt{j}_{1}(kr)\left(P_{f}Q_{i}+Q_{f}P_{i}\right)dr\label{m1-mat}\end{equation}
and\begin{equation}
\left\langle \kappa_{f}\right\Vert m2\left\Vert \kappa_{i}\right\rangle =\frac{30}{\alpha k^{2}}\left\langle j_{f}\right\Vert C_{q}^{(2)}\left\Vert j_{i}\right\rangle \times\int\frac{\kappa_{f}+\kappa_{i}}{3}\mathtt{j}_{2}(kr)\left(P_{f}Q_{i}+Q_{f}P_{i}\right)dr\label{m2-mat}\end{equation}
respectively. Here $j$'s and $\kappa$'s are the total orbital angular
momentum and the relativistic angular momentum quantum numbers respectively;
$k$ is defined as $\omega\alpha$ where $\omega$ is the single particle
difference energy and $\alpha$ is the fine structure constant. The
single particle orbitals are expressed in terms of the Dirac spinors
with $P_{i}$ and $Q_{i}$ as the large and small components for the
$i$th spinor, respectively. The angular coefficients are the reduced
matrix elements of the spherical tensor of rank $m$ and are expressed
as

\begin{equation}
\left\langle \kappa_{f}\right\Vert C_{q}^{(m)}\left\Vert \kappa_{i}\right\rangle =(-1)^{j_{f}+1/2}\sqrt{(2j_{f}+1)(2j_{i}+1)}\left(\begin{array}{ccc}
j_{f} & m & j_{i}\\
\frac{1}{2} & 0 & -\frac{1}{2}\end{array}\right)\pi(l_{f},m,l_{i}),\label{ang-ceoff}\end{equation}
with

\begin{equation}
\pi(l_{f},m,l_{i})=\left\{ \begin{array}{c}
\begin{array}{cc}
1 & \mathrm{if}\: l_{f}+m+l_{i}\,\,\mathrm{even}\\
0 & \mathrm{otherwise}\end{array}\end{array}\right.\label{parity-rules}\end{equation}
 and $l$ being the orbital angular momentum quantum number. When
$z=kr$ is sufficiently small the spherical Bessel function $\mathtt{j}_{n}(z)$
is approximated as 

\begin{equation}
\mathtt{j}_{n}(z)\approx\frac{z^{n}}{(2n+1)!!}=\frac{z^{n}}{1\cdot3\cdot5\cdot\cdot\cdot\cdot(2n+1)}.\label{bessel}\end{equation}

\section{Results and discussions}

\begin{figure}
\begin{centering}
\includegraphics{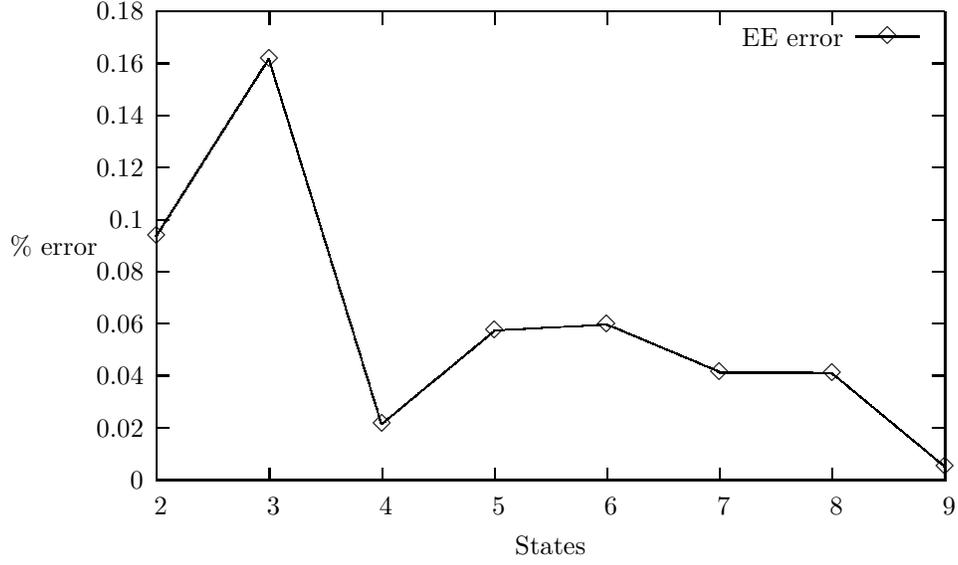}
\par\end{centering}

\caption{\label{ee-error}\% error in calculation of relative transition energies
of O VI. }
\end{figure}

In this manuscript we have studied different electromagnetic transitions
in O VI. We have applied RCCM to calculate the energy levels, line
strengths, transition probabilities and oscillator strengths for the
first nine levels. The calculation of radiative transitions in many-electron
atoms requires a good numerical description of the electronic states
and the wave functions. From the excitation energies presented in
table \ref{en-lev} it is evident that determinations of the energy
levels are very accurate with an average error of less than 0.2\%
which is presented graphically in figure \ref{ee-error}. The comparison
with the presently available theoretical and observed data establishes
the accuracy of RCCM to determine the state energies which are  used
to calculate the transition properties related to astrophysical interests. 

In the actual computation, the Dirac-Fock (DF) ground state and excited
state properties of O VI are computed using the finite basis set expansion
method (FBSE) \cite{rajat-gauss} of a large basis set of $(35s32p25d25f)$
Gaussian functions with the assumption that the nucleus has a Fermi
type finite structure. Excitations from all core orbitals have been
considered within the active space consisting of $14s13p11d10f$ orbitals.
Table \ref{e1-tbl} contains the comparative data for the allowed
electric dipole ($E1$) transitions for all the nine levels. We have
emphasized the two most important fine structure (FS) doublets (($\lambda\lambda$1032,
1038) and ($\lambda\lambda$ 3811, 3834)). The ($\lambda\lambda$1032,
1038) FS doublet is the strongest emission feature in the far-UV spectra
of RR Tel \cite{RRTel} and from the Berkeley spectrograph on ORFEUS
\cite{ORFEUS}. They are also observed in many other symbiotic stars
\cite{tayal}. On the other hand the ($\lambda\lambda$ 3811, 3834)
FS doublet lies in the optical region of many central stars and the
presence of this feature is often referred to as {}``O VI sequence''.
A close comparison of the allowed transition ($E1$) properties with
the GRASP and CIV3 calculations shows that this is the most accurate
determination of the allowed transition probabilities in O VI to date.

\begin{table}

\caption{\label{en-lev}Excitation Energies for different states (in Ryd)
for O VI. RCCM stands for the present calculation. GRASP and CIV3
are the calculations done by Aggarwal \emph{et al} \cite{aggarwal}
and Tayal \cite{tayal} respectively. The numbers presented under
the column NIST are obtained from the database \cite{nist}. }

~

\begin{centering}
\begin{tabular}{clcccc}
\hline 
Index&
States&
RCCM&
NIST &
GRASP  &
CIV3 \tabularnewline
\hline
\hline 
1&
$1s^{2}2s\,^{2}S_{1/2}$&
0.0000&
0.0000&
0.0000&
0.0000\tabularnewline
2&
$1s^{2}2p\, P_{1/2}$&
0.8790&
0.8782&
0.8863&
0.8876\tabularnewline
3&
$1s^{2}2p\,^{2}P_{3/2}$&
0.8845&
0.8831&
0.8910&
0.8944\tabularnewline
4&
$1s^{2}3s\,^{2}S_{1/2}$&
5.8337&
5.8325&
5.8251&
5.8292\tabularnewline
5&
$1s^{2}3p\,^{2}P_{1/2}$&
6.0735&
6.0701&
6.0646&
6.0676\tabularnewline
6&
$1s^{2}3p\,^{2}P_{3/2}$&
6.0751&
6.0715&
6.0660&
6.0696\tabularnewline
7&
$1s^{2}3d\,^{2}D_{3/2}$&
6.1502&
6.1476&
6.1391&
6.1421\tabularnewline
8&
$1s^{2}3d\,^{2}D_{5/2}$&
6.1506&
6.1481&
6.1395&
6.1427\tabularnewline
9&
$1s^{2}4s\,^{2}S_{1/2}$&
7.7732&
7.7703&
7.7615&
7.7663\tabularnewline
\hline
\end{tabular}
\par\end{centering}
\end{table}

\begin{table}

\caption{\label{e1-tbl}Upper (lower) states $f$($i$), transition wavelengths
($\lambda_{fi}$) in $\textrm{\AA}$, allowed (electric dipole) line-strengths
($\mathtt{S}_{fi}$), transition probabilities ($\mathtt{A}_{fi}$
in $\mathrm{sec^{-1}}$) and oscillator strengths ($\mathtt{f}_{fi}$)
in O VI. Numbers ($x$) in parentheses represent $10^{x}$. RCCM labels
the present calculation. The present calculation agrees well with
the MCHF calculations done by Froese Fischer \emph{et al} \cite{mchf}.}

~

\begin{centering}
\begin{tabular}{cc|cc|llll|llcl}
\hline 
$f$&
$i$&
$\lambda_{fi}$ &
$\mathtt{S}_{fi}$&
\multicolumn{4}{c|}{$\mathtt{A}_{fi}$}&
\multicolumn{4}{c}{$\mathtt{f}_{fi}$}\tabularnewline
\hline
\hline 
&
&
&
RCCM&
RCCM &
NIST&
GRASP&
CIV3&
RCCM&
NIST&
GRASP&
CIV3\tabularnewline
2&
1&
1037.63&
0.4502&
4.0824(8)&
4.09(8)&
4.2487(8)&
4.33(8)&
6.5896(-2)&
6.60(-2)&
6.7338(-2)&
6.77(-2)\tabularnewline
3&
1&
1031.92&
0.9014&
4.1549(8)&
4.16(8)&
4.3212(8)&
4.39(8)&
1.3265(-1)&
1.33(-1)&
1.3552(-1)&
1.36(-1)\tabularnewline
4&
2&
183.94&
0.0348&
5.6640(9)&
5.70(9)&
5.6283(9)&
5.65(9)&
2.8729(-2)&
2.89(-2)&
2.8726(-2)&
2.89(-2)\tabularnewline
4&
3&
184.12&
0.0700&
1.1361(10)&
1.14(10)&
1.1280(10)&
1.13(10)&
2.8870(-2)&
2.89(-2)&
2.8841(-2)&
2.90(-2)\tabularnewline
5&
1&
150.12&
0.0861&
2.5796(10)&
2.62(10)&
2.5791(10)&
2.60(10)&
8.7158(-2)&
8.74(-2)&
8.7299(-2)&
8.79(-2)\tabularnewline
5&
4&
3835.30&
2.8158&
5.0564(7)&
5.05(7)&
5.1726(7)&
5.14(7)&
1.1150(-1)&
1.11(-1)&
1.1224(-1)&
1.12(-1)\tabularnewline
6&
1&
150.09&
0.1717&
2.5717(10)&
2.62(10)&
2.5716(10)&
2.60(10)&
1.7370(-1)&
1.77(-1)&
1.7401(-1)&
1.75(-1)\tabularnewline
6&
4&
3812.52&
5.6347&
5.1503(7)&
5.14(7)&
5.2664(7)&
5.23(7)&
2.2446(-1)&
2.24(-1)&
2.2590(-1)&
2.25(-1)\tabularnewline
7&
2&
172.93&
0.7451&
7.2975(10)&
7.33(10)&
7.3084(10)&
7.29(10)&
6.5437(-1)&
6.58(-1)&
6.5950(-1)&
6.58(-1)\tabularnewline
7&
3&
173.09&
0.1493&
1.4582(10)&
1.46(10)&
1.4604(10)&
1.46(10)&
6.5501(-2)&
6.57(-2)&
6.6009(-2)&
6.59(-2)\tabularnewline
7&
5&
11746.17&
3.8098&
1.1907(6)&
1.19(6)&
1.0478(6)&
1.01(6)&
4.9259(-2)&
4.90(-2)&
4.7036(-2)&
4.64(-2)\tabularnewline
7&
6&
11965.18&
0.7618&
2.2252(5)&
2.24(5)&
1.9794(5)&
1.91(5)&
4.8347(-3)&
4.81(-3)&
4.6144(-3)&
4.55(-3)\tabularnewline
8&
3&
172.99&
1.3439&
8.7667(10)&
8.78(10)&
8.7623(10)&
8.74(10)&
5.8994(-1)&
5.91(-1)&
5.9400(-1)&
5.93(-1)\tabularnewline
8&
6&
11894.88&
6.8581&
1.5331(6)&
1.37(6)&
1.2087(6)&
1.17(6)&
4.5388(-2)&
4.36(-2)&
4.1782(-2)&
4.12(-2)\tabularnewline
9&
2&
132.22&
0.0051&
2.2174(9)&
2.18(9)&
2.1448(9)&
2.18(9)&
5.8116(-3)&
5.70(-3)&
5.6488(-3)&
5.75(-3)\tabularnewline
9&
3&
132.31&
0.0102&
4.4464(9)&
4.34(9)&
4.2946(9)&
4.40(9)&
5.8349(-3)&
5.70(-3)&
5.6633(-3)&
5.80(-3)\tabularnewline
9&
5&
535.96&
0.2289&
1.5059(9)&
1.48(9)&
1.4819(9)&
1.51(9)&
6.4851(-2)&
6.39(-2)&
6.4073(-2)&
6.50(-2)\tabularnewline
9&
6&
536.41&
0.4725&
3.1011(9)&
2.96(9)&
2.9700(9)&
3.02(9)&
6.6885(-2)&
6.39(-2)&
6.4312(-2)&
6.53(-2)\tabularnewline
\hline
\end{tabular}
\par\end{centering}
\end{table}

Tables \ref{e2-tbl}-\ref{m2-tbl} contain the details of the forbidden
transitions in O VI. This is the first theoretical determination of
forbidden transition probabilities and oscillator strengths for O
VI. They are related to the corresponding line strengths by Eqs. (\ref{E2}-\ref{M2}).
From the data presented in tables (\ref{e1-tbl}-\ref{m2-tbl}) we
have determined the radiative lifetimes of the excited states which
decay via allowed and forbidden transitions to lower levels. We have
observed that among the forbidden transitions between the three fine
structure doublets, the $E2$ line-strengths ($\mathtt{S}_{E2}$)
are greater than $M1$ line-strengths ($\mathtt{S}_{M1}$) except
for $1s^{2}2p\,^{2}P_{3/2}\longrightarrow1s^{2}2p\,^{2}P_{1/2}$ transition
for which the wavelength lies in the near infrared (IR) region, whereas
the other two are in the IR ($1s^{2}3p\,^{2}P_{3/2}\longrightarrow1s^{2}3p\,^{2}P_{1/2}$
) and far from the IR region ($1s^{2}3d\,^{2}D_{5/2}\longrightarrow1s^{2}3d\,^{2}D_{3/2}$).
These values are given in tables \ref{e2-tbl} and \ref{m1-tbl}. 

We have shown how the electric quadrupole ($E2$) and magnetic dipole
($M1$) transition probabilities depend on the transition wavelengths
in figure \ref{a-value} and \ref{f-value}. The abscissas in those
two figures represent the different transitions involved. For example
$3\longrightarrow2$ means transition from level $3$ ($1s^{2}2p\,^{2}P_{3/2}$)
to level $2$ ($1s^{2}2p\,^{2}P_{1/2}$). The corresponding wavelengths
are given in table \ref{e2-tbl} and \ref{m1-tbl}. On the other hand
the ordinates in figure \ref{a-value} and \ref{f-value} correspond
to the logarithms of $\mathtt{A}$ and $\mathtt{f}$ values respectively.
From these two figures we have observed that for transitions between
two fine structure levels, the $M1$ transition probabilities and
hence the oscillator strengths are greater than those of the corresponding
$E2$ transitions. Whereas in the other cases the trend is the opposite.
This feature can be analyzed by an order of magnitude estimation following
Landau and Lifshitz \cite{landau}. as given in the next subsection
(\ref{order-mag}). 

\begin{table}

\caption{\label{e2-tbl}Upper (lower) states $f$($i$), transition wavelengths
($\lambda_{fi}$) in $\textrm{\AA}$, electric quadrupole ($E2$)
line-strengths ($\mathtt{S}_{fi}$), transition probabilities ($\mathtt{A}_{fi}$
in $\mathrm{sec^{-1}}$) and oscillator strengths ($\mathtt{f}_{fi}$)
in O VI. Numbers ($x$) in parentheses represent $10^{x}$. }

~

\begin{centering}
\begin{tabular}{cc|rl|ll}
\hline 
$f$&
$i$&
$\lambda_{fi}$ &
$\mathtt{S}_{fi}$&
$\mathtt{A}_{fi}$&
$\mathtt{f}_{fi}$\tabularnewline
\hline
3&
2&
187503.70&
0.4838&
5.8446(-10)&
6.1609(-15)\tabularnewline
5&
3&
175.68&
0.3540&
1.1845(6)&
2.7404(-6)\tabularnewline
6&
2&
175.47&
0.3519&
5.9233(5)&
5.4683(-6)\tabularnewline
6&
3&
175.63&
0.3615&
6.0557(5)&
2.8005(-6)\tabularnewline
6&
5&
641738.03&
18.6557&
4.7989(-11)&
5.9256(-15)\tabularnewline
7&
1&
148.23&
1.0537&
4.1222(6)&
2.7157(-5)\tabularnewline
7&
4&
2891.26&
12.0391&
1.6683(1)&
4.1814(-8)\tabularnewline
8&
1&
148.15&
1.5807&
4.1336(6)&
4.0805(-5)\tabularnewline
8&
4&
2861.66&
18.0659&
1.7571(1)&
6.4714(-8)\tabularnewline
8&
7&
279530.06&
4.2353&
4.6320(-10)&
8.1388(-15)\tabularnewline
9&
7&
561.58&
2.2936&
2.3996(4)&
5.6725(-7)\tabularnewline
9&
8&
562.71&
3.5945&
3.5674(4)&
5.6449(-7)\tabularnewline
\hline
\end{tabular}
\par\end{centering}
\end{table}

\begin{table}

\caption{\label{m1-tbl}Upper (lower) states $f$($i$), transition wavelengths
($\lambda_{fi}$) in $\textrm{\AA}$, magnetic dipole ($M1$) line-strengths
($\mathtt{S}_{fi}$), transition probabilities ($\mathtt{A}_{fi}$
in $\mathrm{sec^{-1}}$) and oscillator strengths ($\mathtt{f}_{fi}$)
in O VI. Numbers($x$) in parentheses represent $10^{x}$. }

~

\begin{centering}
\begin{tabular}{cc|rl|ll}
\hline 
$f$&
$i$&
$\lambda_{fi}$ &
$\mathtt{S}_{fi}$&
$\mathtt{A}_{fi}$&
$\mathtt{f}_{fi}$\tabularnewline
\hline
3&
2&
187503.70&
1.3333&
1.3639(-3)&
1.4378(-8)\tabularnewline
4&
1&
156.24&
5.808(-6)&
2.0538(1)&
7.5166(-11)\tabularnewline
5&
2&
175.52&
2.1609(-6)&
5.3898&
2.4894(-11)\tabularnewline
5&
3&
175.68&
6.2410(-7)&
1.5523&
3.5915(-12)\tabularnewline
6&
2&
175.47&
2.9160(-7)&
3.6396(-1)&
3.3602(-12)\tabularnewline
6&
3&
175.63&
7.6038(-5)&
9.4641(1)&
4.3769(-10)\tabularnewline
6&
5&
641738.03&
1.3377&
3.4133(-5)&
4.2148(-9)\tabularnewline
7&
1&
148.23&
1.9600(-8)&
4.0581(-2)&
2.6736(-13)\tabularnewline
7&
4&
2891.26&
4.0000(-10)&
1.1160(-7)&
2.7974(-16)\tabularnewline
8&
7&
279530.06&
2.3997&
4.9393(-4)&
8.6792(-9)\tabularnewline
9&
1&
117.27&
1.4400(-6)&
1.2041(1)&
2.4828(-11)\tabularnewline
9&
4&
470.24&
5.5225(-6)&
7.1627(-1)&
2.3746(-11)\tabularnewline
9&
7&
561.58&
1.0000(-10)&
7.6151(-6)&
1.8003(-16)\tabularnewline
\hline
\end{tabular}
\par\end{centering}
\end{table}

\begin{table}

\caption{\label{m2-tbl}Upper (lower) states $f$($i$), transition wavelengths
($\lambda_{fi}$) in $\textrm{\AA}$, magnetic quadrupole ($M2$)
line-strengths ($\mathtt{S}_{fi}$), transition probabilities ($\mathtt{A}_{fi}$
in $\mathrm{sec^{-1}}$) and oscillator strengths ($\mathtt{f}_{fi}$)
in O VI. Numbers ($x$) in parentheses represent $10^{x}$. }

~

\begin{centering}
\begin{tabular}{cc|rl|ll}
\hline 
$f$&
$i$&
$\lambda_{fi}$ &
$\mathtt{S}_{fi}$&
$\mathtt{A}_{fi}$&
$\mathtt{f}_{fi}$\tabularnewline
\hline
3&
1&
1031.92&
6.8956(-3)&
2.1966(-5)&
7.0158(-15)\tabularnewline
4&
3&
184.12&
2.687(-2)&
9.4686(-1)&
2.4068(-12)\tabularnewline
6&
1&
150.01&
1.9721(1)&
9.6513(2)&
6.5210(-9)\tabularnewline
6&
4&
3812.52&
2.0035(1)&
9.2715(-5)&
4.0420(-13)\tabularnewline
7&
2&
172.93&
2.7019(-3)&
6.5114(-2)&
5.8407(-13)\tabularnewline
7&
3&
173.09&
8.3694(-4)&
2.0077(-2)&
9.0210(-14)\tabularnewline
7&
5&
11746.17&
3.2436(-4)&
5.4071(-12)&
2.2376(-19)\tabularnewline
7&
6&
11965.18&
1.7108(-4)&
2.6004(-12)&
5.5830(-20)\tabularnewline
8&
2&
172.83&
2.4806(-4)&
3.9977(-3)&
5.3723(-14)\tabularnewline
8&
3&
172.99&
5.4612(-5)&
8.7607(-4)&
5.8973(-15)\tabularnewline
8&
6&
11474.04&
3.8025(-6)&
4.7513(-14)&
1.4071(-21)\tabularnewline
9&
3&
132.31&
8.4971(-3)&
1.5621(0)&
2.0506(-12)\tabularnewline
9&
6&
536.41&
2.0101(1)&
3.3745(0)&
7.2804(-11)\tabularnewline
\hline
\end{tabular}
\par\end{centering}
\end{table}

Table \ref{lifetime} presents the comprehensive values of the lifetimes
which are compared with the available theoretical data determined
by using the non-relativistic multi-configuration Hartree-Fock (MCHF)
calculations by Charlotte F. Fischer's group \cite{mchf}. We have
calculated all possible decay rates of the excited states presented
in figure \ref{decay-channels}. Our calculated lifetimes are in good
agreement with Froese Fischer's data except for the $1s^{2}4s\,^{2}S_{1/2}$
state which is dominated by the electric dipole allowed ($E1$) decay
to the $1s^{2}2p\,^{2}P_{3/2}$ state (given by $9\longrightarrow3$
decay in table \ref{e1-mat}). In the present calculation we have
considered all possible excitations of the core electrons to the virtual
apace. This leads to a complete treatment of correlation effects in
a more elegant and compact way. Higher order excitations are also
taken into account through perturbative triple excitations which we
have applied earlier and can be found in Refs. \cite{ccsd(t),kaldor-book}.
Inclusion of higher order multipoles leads to the most accurate transition
probability data sets of O VI to our knowledge.

\begin{table}

\caption{\label{lifetime}Lifetimes of O VI states. MCHF labels Multi-configuration
Hartree-Fock calculations \cite{mchf}. Numbers ($x$) in parentheses
represent $10^{x}$.}

~

\begin{centering}
\begin{tabular}{l|rr}
\hline 
States&
\multicolumn{2}{r}{Lifetimes (in $\mathrm{sec^{-1}}$)}\tabularnewline
\hline
\hline 
&
RCCM&
MCHF\tabularnewline
$1s^{2}2p\, P_{1/2}$&
2.44 95(-9)&
2.44 60(-9)\tabularnewline
$1s^{2}2p\,^{2}P_{3/2}$&
2.40 68(-9)&
2.40 43(-9)\tabularnewline
$1s^{2}3s\,^{2}S_{1/2}$&
5.87 37(-11)&
5.85 46(-11)\tabularnewline
$1s^{2}3p\,^{2}P_{1/2}$&
3.86 88(-11)&
3.78 54(-11)\tabularnewline
&
&
\tabularnewline
$1s^{2}3p\,^{2}P_{3/2}$&
3.88 05(-11)&
3.81 14(-11)\tabularnewline
$1s^{2}3d\,^{2}D_{3/2}$&
1.14 20(-11)&
1.14 09(-11)\tabularnewline
$1s^{2}3d\,^{2}D_{5/2}$&
1.14 06(-11)&
1.14 23(-11)\tabularnewline
$1s^{2}4s\,^{2}S_{1/2}$&
11.04 55(-11)&
9.06 23(-11)\tabularnewline
\hline
\end{tabular}
\par\end{centering}
\end{table}

Branching ratio of any decay of a particular state depends on the
decay probability of that state to all the lower states and is defined
as

\begin{equation}
\Gamma_{\mathrm{up}\longrightarrow\mathrm{low}}=\frac{A_{\mathrm{up}\longrightarrow\mathrm{low}}}{{\displaystyle \sum_{\mathrm{low}}}A_{\mathrm{up}\longrightarrow\mathrm{low}}}\,.\label{Br-ratio}\end{equation}
We have analyzed the branching ratio of the $1s^{2}4s\,^{2}S_{1/2}$
state to different lower states which decay dominantly via electric
dipole ($E1$) transition. The transition rates are given in table
\ref{e1-tbl}. Table \ref{BR-table} contains our calculated value
of $\Gamma_{\mathrm{up}\longrightarrow\mathrm{low}}$ and the comparison
with other available data. In the present calculation, the denominator
in Eq. (\ref{Br-ratio}) consists of the decay rates of the upper
state ($1s^{2}4s\,^{2}S_{1/2}$) to the lower states via all possible
allowed and forbidden channels, whereas in the other theoretical calculations
only allowed transitions have been considered. To show the accuracy
of our calculation we have presented the results upto sixth decimal
places. Our calculated values of the branching ratios are off by 0.3\%
- 1.9\% as compared to the observed value \cite{nist}. Among these
transitions, the particular decay ($(\mathrm{up})1s^{2}4s\,^{2}S_{1/2}\longrightarrow(\mathrm{low})1s^{2}2p\,^{2}P_{3/2}$)
is the dominant one which is within an error of 0.3\%. Similarly the
branching ratios of decay of the other states can be easily obtained
from this work and will be very useful for astrophysical and related
studies.

\begin{table}

\caption{\label{BR-table}Branching ratios of $(\mathrm{up})1s^{2}4s\,^{2}S_{1/2}\longrightarrow(\mathrm{low})1s^{2}2p\,^{2}P_{3/2}$
, $(\mathrm{up})1s^{2}4s\,^{2}S_{1/2}\longrightarrow(\mathrm{low})1s^{2}2p\,^{2}P_{1/2}$,
$(\mathrm{up})1s^{2}4s\,^{2}S_{1/2}\longrightarrow(\mathrm{low})1s^{2}3p\,^{2}P_{3/2}$
and $(\mathrm{up})1s^{2}4s\,^{2}S_{1/2}\longrightarrow(\mathrm{low})1s^{2}3p\,^{2}P_{1/2}$
transitions. The decays rates of the $1s^{2}4s\,^{2}S_{1/2}$ states
are dominated by electric dipole ($E1$) transition. Each row corresponds
to the value of the branching ratios of the upper level correspond
to different $E1$ transitions obtained by different methods. The
values given under NIST are obtained from the NIST database \cite{nist}.}

~

\begin{centering}
\begin{tabular}{l|c|c|c|c}
\hline 
Methods&
$\Gamma_{1s^{2}4s\,^{2}S_{1/2}\longrightarrow1s^{2}2p\,^{2}P_{3/2}}$&
$\Gamma_{1s^{2}4s\,^{2}S_{1/2}\longrightarrow1s^{2}2p\,^{2}P_{1/2}}$&
$\Gamma_{1s^{2}4s\,^{2}S_{1/2}\longrightarrow1s^{2}3p\,^{2}P_{3/2}}$&
$\Gamma_{1s^{2}4s\,^{2}S_{1/2}\longrightarrow1s^{2}3p\,^{2}P_{1/2}}$\tabularnewline
\hline
\hline 
RCCM&
0.394 548&
0.196 737&
0.275 143&
0.133 610\tabularnewline
MCHF&
0.398 581&
0.199 117&
0.267 757&
0.134 543\tabularnewline
GRASP&
0.394 314&
0.196 928&
0.272 695&
0.136 063\tabularnewline
CIV3&
0.396 040&
0.196 220&
0.271 827&
0.135 914\tabularnewline
NIST&
0.395 985&
0.198 905&
0.270 073&
0.135 036\tabularnewline
\hline
\end{tabular}
\par\end{centering}
\end{table}

\subsection{\label{order-mag}Order of magnitude estimation }

\begin{figure}
\begin{centering}
\includegraphics{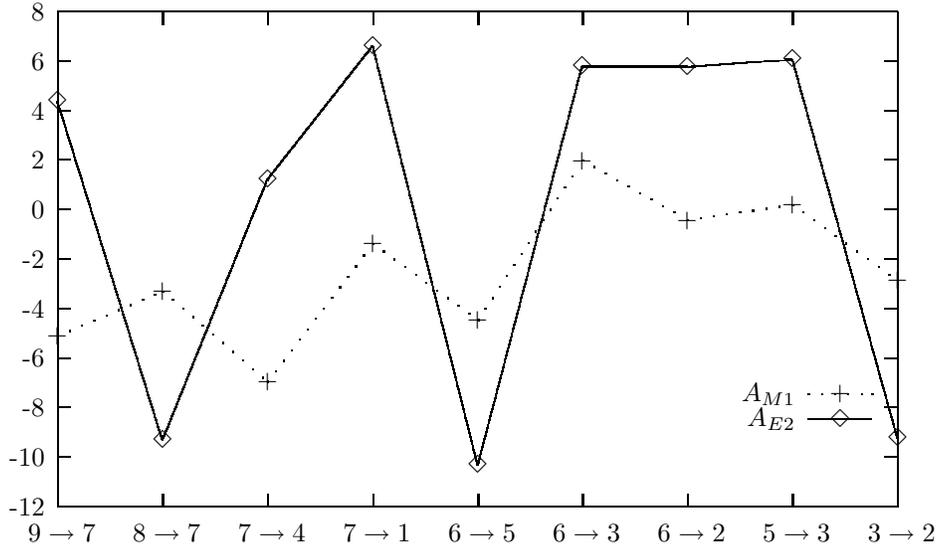}
\par\end{centering}

\caption{\label{a-value}Dependencies of transition probability ($\mathtt{A}$)
on energies. The abscissa represent the energies for different transitions
and the ordinate is the values of the $\mathtt{A}$ values in a log
scale. Here $3\longrightarrow2$ means transition from level $3$
($1s^{2}2p\,^{2}P_{3/2}$) to level $2$ ($1s^{2}2p\,^{2}P_{1/2}$).
The levels $1-9$ are identified as $1s^{2}2s\,^{2}S_{1/2}$, $1s^{2}2p\, P_{1/2}$,
$1s^{2}2p\,^{2}P_{3/2}$, $1s^{2}3s\,^{2}S_{1/2}$, $1s^{2}3p\,^{2}P_{1/2}$,
$1s^{2}3p\,^{2}P_{3/2}$, $1s^{2}3d\,^{2}D_{3/2}$, $1s^{2}3d\,^{2}D_{5/2}$
and $1s^{2}4s\,^{2}S_{1/2}$ respectively as denoted in table \ref{en-lev}
in the column `Index'. }
\end{figure}

\begin{figure}
\begin{centering}
\includegraphics{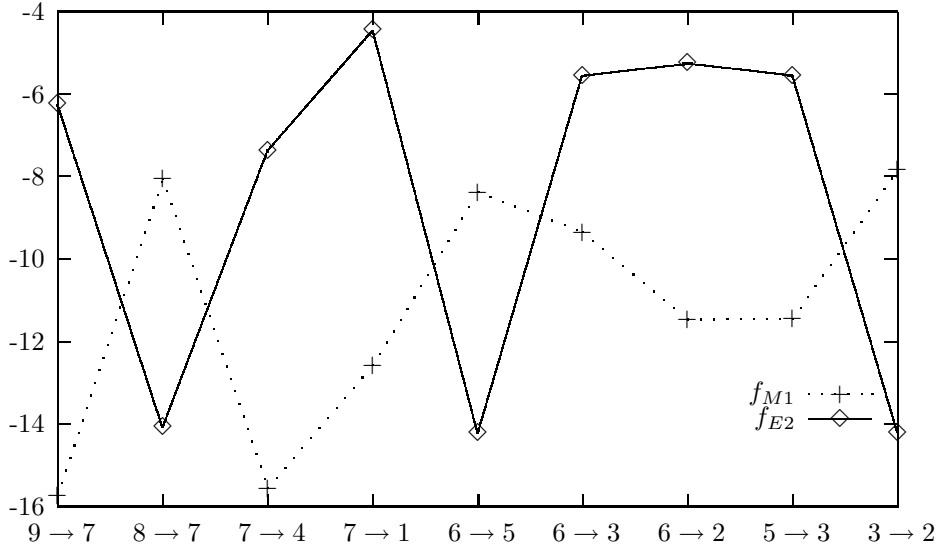}
\par\end{centering}

\caption{\label{f-value}Dependencies of oscillator strengths ($\mathtt{f}$)
on transition energies. The abscissa represent the energies for different
transitions and the ordinate is the values of the $\mathtt{f}$ values
in a log scale. Identification of different levels are same as described
in Fig. (\ref{a-value}).}
\end{figure}

Following Landau and Lifshitz \cite{landau} the magnetic moment of
an atom is equal in order of magnitude to the Bohr magneton $\mu\sim\frac{e\hbar}{mc}$
with all the variables with their usual meaning. This differs by a
factor of $\alpha$, the fine structure constant, from the order of
magnitude of the electric dipole moment $d\sim ea_{0}\sim\hbar^{2}/me$.
Since $v/c\sim\alpha$, we have approximated $\mu\sim dv/c$. Hence
it follows that the probability of $M1$ radiation from the atom is
about $\alpha^{2}$ times less than the electric dipole ($E1$) radiation
at the same frequency. Magnetic radiation is therefore important in
practice only for forbidden transitions governed by the selection
rules for the electric case. The ratio of $E2$ to that of $M1$ radiation
in order of magnitude for a given transition frequency $\omega$ is 

\begin{equation}
\frac{E2}{M1}\sim\frac{(ea_{0})^{2}\omega^{2}/c^{2}}{\mu^{2}}\sim\frac{a_{0}^{4}m^{2}\omega^{2}}{\hbar^{2}}\sim\left(\frac{\Delta E}{E}\right)^{2}\label{magnitude}\end{equation}
where $\Delta E$ is the change in energy in the transition involved
and the order of magnitude of the quadrupole moment and the energy
of the atom ($E$) are $\sim ea_{0}^{2}$ and $\sim\hbar^{2}/ma_{0}^{2}$
respectively. For medium atomic transition frequencies ($\Delta E\sim E$)
the probabilities of $E2$ and $M1$ are of the same order of magnitude.
If, however, $\Delta E\ll E$, \emph{i.e.} for transitions between
two fine structure components of the same term the $M1$ transition
probability is more probable than that of $E2$. The results presented
in table \ref{e2-tbl} and \ref{m1-tbl} and in the figures \ref{a-value}
and \ref{f-value} follow the pattern. 

\section{Conclusion}

In this paper we have performed relativistic coupled-cluster calculations
of all possible allowed and forbidden radiative transition probabilities
of the first eight excited states of O VI to the ground state. The
lifetimes of those excited states have been determined considering
all possible decay channels. This paper reports the first and most
comprehensive calculation of the forbidden transition probabilities
to our knowledge. Comparison with other available data are made for
allowed transitions and radiative lifetimes. Accuracy of the computed
transition data depends on both the accuracy of the state energies
(transition energies) and the line strengths. Hence the determination
of the state energies with an error of (and less than) 2\% are used
to benchmark the RCCM theory. We also made graphical analysis of the
energy dependencies of the $E2$ and $M1$ transition probabilities
and discussed the trends of different transitions with order of magnitude
estimates. We believe that the present data sets will be useful for
astrophysical studies of stellar spectra.

\begin{verse}
\textbf{Acknowledgment} : These computations are carried out in the
Intel Xeon cluster at the Department of Astronomy, OSU under the Cluster-Ohio
initiative. This work was partially supported by the National Science
Foundation and the Ohio State University (CS). RKC acknowledges the
Department of Science and Technology, India (grant SR/S1/PC-32/2005).
We gratefully acknowledge Prof. Russell Pitzer and Prof. Anil Pradhan
for their comments and criticism on the manuscript.
\end{verse}

\end{document}